\documentclass[preprint2]{aastex}
\usepackage{spr-astr-addons}
\usepackage{graphicx}
\newcommand{\ltapprox}{\raisebox{-0.5ex}{$\,\stackrel{<}{\scriptstyle
\sim}\,$}}

\newcommand{\ri}{r_{\rm i}}
\newcommand{\rw}{r_{\rm w}}
\newcommand{\rg}{r_{\rm g}}
\newcommand{\phiG}{\phi_{\scriptscriptstyle \rm G}}
\newcommand{\cs}{c_{\rm s}}

\newcommand{\OmegaK}{\Omega_{\scriptscriptstyle \rm K}}
\newcommand{\fa}{f_{\rm a}}
\newcommand{\fw}{f_{\rm w}}
\newcommand{\Fd}{F_{\rm d}}
\newcommand{\Ld}{L_{\rm d}}
\newcommand{\LEdd}{L_{\rm Edd}}
\newcommand{\Pa}{P_{\rm a}}
\newcommand{\Pw}{P_{\rm w}}
\newcommand{\sigmaT}{\sigma_{\scriptscriptstyle \rm T}}
\newcommand{\Mdot}{\dot M}
\newcommand{\Mdota}{\dot M_{\rm a}}
\newcommand{\Mdotw}{\dot M_{\rm w}}
\newcommand{\ddr}{\frac{\rm d}{{\rm d}r}}
\newcommand{\dr}{{\rm d}r}
\newcommand{\dz}{{\rm d}z}

\begin{document}
\title{Towards a New Standard Model for Black Hole Accretion}

\shorttitle{Black Hole Accretion}        
\shortauthors{Kuncic \& Bicknell}

\author{Zdenka Kuncic}
\affil{School of Physics, University of Sydney, Sydney, NSW, Australia
2006}

\author{Geoffrey V. Bicknell}
\affil{Research School of Astronomy \& Astrophysics, Australian National University, Cotter Rd., Weston ACT, Australia 2611}





\begin{abstract}
We briefly review recent developments in black hole accretion disk
theory, emphasizing the vital role played by magnetohydrodynamic (MHD)
stresses in transporting angular momentum. The apparent universality
of accretion-related outflow phenomena is a strong indicator that
large-scale MHD torques facilitate vertical transport of angular
momentum. This leads to an enhanced overall rate of angular momentum
transport and allows accretion of matter to proceed at an interesting
rate. Furthermore, we argue that when vertical transport is important,
the radial structure of the accretion disk is modified at small
radii and this affects the disk emission spectrum. We present a simple
model demonstrating how energetic, magnetically-driven outflows modify the
emergent disk emission spectrum with respect to that predicted by
standard accretion disk theory. A comparison of the predicted spectra
against observations of quasar spectral energy distributions suggests
that mass accretion rates inferred using the standard disk model may
be severely underestimated.
\end{abstract}


\keywords{accretion, accretion disks --- black hole physics ---
  (magnetohydrodynamics:) MHD --- radiation mechanisms: general ---
  turbulence waves --- galaxies: active --- X-rays: binaries}

\section{Introduction}
\label{intro}

The standard theory of astrophysical disk accretion was formulated over
thirty years ago \citep{prinrees72,novthorn73,SS73}.
Since then, arguably the most important theoretical milestone has been
the demonstration via numerical simulations that the removal of angular
momentum required for accretion to proceed is mediated by 
magnetohydrodynamic (MHD) turbulence driven by the weak-field
magnetorotational instability (MRI -- see \citealt{balbus03} for a review).
Notwithstanding these groundbreaking developments, it still remains
unclear precisely how energy can be channelled from the bulk accretion flow
to a diffuse region outside the disk where high-energy processes can operate.
This outstanding issue is inextricably linked with the formation
and ubiquity of outflow phenomena associated with accretion disks.
Indeed, the formation and nature of relativistic jets remains one of the
most formidable theoretical problems in this field.

In this paper, we highlight some recent theoretical progress made towards
understanding vertical transport in accretion disks \citep{kunbick04}.
A macroscopic, mean field approach is adopted for the MHD turbulence.
This allows us to identify the dominant mechanisms responsible for
angular momentum transport as well as the main contributions to the global
energy budget of the system. We also demonstrate quantitatively how outflows, both
Poynting-flux-dominated and mass-flux-dominated, can modify the
spectrum of disk emission substantially from that predicted by
standard accretion disk theory.


\section{MHD disk accretion}
\label{s:MHD}

Here, we present statistically-averaged equations in a
cylindrical $(r,\phi,z)$ coordinate system for a fluid which is
time-independent and axisymmetric in the mean
($\langle \partial / \partial t \rangle = \langle \partial /\partial
\phi \rangle =0$, where the angle brackets denote ensemble-averaged quantities).
We use Newtonian physics throughout, with the gravitational
potential $\phiG = -GM(r^2 + z^2)^{-1/2}$.
The fluid description is valid down to the innermost marginally stable 
orbit $\ri \approx 6\rg$, where $\rg = GM/c^2$ is the gravitational
radius of a black hole of mass $M$.
The mean-field conservation equations are integrated vertically over an
arbitrary disk scaleheight, $h = h(r)$.
Quantities calculated at the disk surface ($z= \pm h$) are denoted by a $\pm$
superscript, $X^{\pm}$, and we assume reflection symmetry about the disk midplane,
so that $|X^+| = |X^-|$.
Midplane values of variables are denoted by $X_0$.
We assume $h$ is much less than the radius (the ``thin disk''
approximation), so that quantities of order $h/r$ and ${\rm d}h/{\rm d}r$ are neglected.

We adopt a mass-weighted statistical averaging approach in 
which all variables are decomposed into mean and fluctuating parts,
with intensive variables such as the velocity mass averaged according
to
\begin{equation}
v_i = \tilde v_i + v_i^\prime \qquad , \qquad \langle \rho v_i^\prime
\rangle =0 \qquad ,
\end{equation}
while extensive variables, such as density, pressure and magnetic
field, averaged the following way:
\begin{eqnarray}
\rho = \bar \rho + \rho^\prime \> &,& \> \langle \rho^\prime
\rangle =0 \qquad ,  \\
p = \bar p + p^\prime \> &,& \>  \langle p^\prime
\rangle =0 \qquad , \\
B_i = \bar B_i + B_i^\prime \> &,&  \> \langle B_i^\prime
\rangle =0 \qquad .
\end{eqnarray}
Note that intensive averages are denoted by a tilde, extensive
averages are denoted by a bar and fluctuating components are denoted
by a prime. The fluctuating velocity components are restricted
to subsonic speeds because the MRI is a weak-field instability which
drives subsonic turbulence with $ \langle \rho v^{\prime 2} \rangle
\ltapprox \langle \rho c_{\rm s}^2 \rangle \ll \rho v^2 \simeq \rho
r^2 {\OmegaK}^2$, where
$c_{\rm s} = (kT/\mu m_{\rm p})^{1/2}$
is the local sound speed and $\OmegaK = (GM/r^3)^{1/2}$ is the keplerian angular
velocity. The only restriction we place on the mean fluid velocity
components is that they
satisfy $\tilde v_\phi \gg \tilde v_r ,  \tilde v_z$ and that $\tilde
v_r$ and $\tilde v_\phi$ are
independent of $z$, simplifying the vertical integration.
The combined magnetic and Reynolds stresses are defined by
\begin{equation}
\langle t_{ij} \rangle = \langle \frac{B_i B_j}{4\pi} \rangle -
\delta_{ij} \langle \frac{B^2}{8\pi} \rangle - \langle \rho v_i^\prime v_j^\prime \rangle
\label{e:tij}
\end{equation}

In what follows, we summarize the salient equations for radial and
vertical transport of mass, angular momentum and energy. A full
derivation of these equations can be found in our earlier
paper \citep{kunbick04}. For clarity, we have simplified the
presentation of the relevant equations by omitting negligible
correlation terms in the energy equation\footnote{In particular, triple
  correlation terms of the form $\langle t_{ij} v_j^\prime \rangle$
  are negligible compared to analogous correlations with the mean
  fluid velocity $\langle t_{ij} \rangle \tilde v_j $.} and by removing the notation for averaged extensive
and intensive quantities. Thus, averaged quantities are implicitly assumed.

\subsection{Mass transfer}
\label{s:mass}

Vertical integration of the mean-field continuity equation gives
\begin{equation}
\ddr \int_{-h}^{+h} 2 \pi r  \rho  v_r \> \dz
\, + \, 4 \pi r  \rho^+  v_z^+  = 0  \qquad .
\label{e:mass_int}
\end{equation}
We now introduce the usual definitions for the surface mass density,
\begin{equation}
\Sigma(r) \equiv \int_{-h}^{+h}  \rho \> \dz
\label{e:sigma_defn}
\end{equation}
and mass accretion rate,
\begin{equation}
\Mdota (r) \equiv 2\pi r  \Sigma (-  v_r)  \qquad .
\label{e:Mdota}
\end{equation}
We also introduce an analogous mass outflow rate,
\begin{equation}
\Mdotw (r) = \Mdotw (\ri) - \int_{\ri}^{r} 4\pi r  \rho^+  v_z ^+{\rm d}r
\label{e:Mdotw}
\end{equation}
associated with a mean vertical velocity $ v_z^+$ at the disk 
surface, i.e. at the base of a disk wind.
In terms of the above definitions, the vertically integrated continuity equation,
(\ref{e:mass_int}), can be written as
\begin{equation}
\ddr \Mdota (r) = 4\pi r  \rho^+ v_z ^+ = - \ddr \Mdotw (r)
\label{e:dM}
\end{equation}
implying that
\begin{equation}
\Mdota (r) + \Mdotw (r) = \Mdota (\ri) + \Mdotw (\ri) = \hbox{constant} = \Mdot
\label{e:Mdot}
\end{equation}
where $\Mdota (\ri) + \Mdotw (\ri)$ is the total mass flux at the innermost
stable orbit, $\ri$.
Equation (\ref{e:dM}) implies that under steady--state conditions,
the radial mass inflow decreases towards small $r$ at the same rate as the
vertical mass outflow increases in order to maintain a constant nett mass flux,
$\Mdot$, which is the nett accretion rate at $r =
\infty$, i.e. $\Mdot = \Mdota (\infty)$.

\subsection{Angular momentum}
\label{s:ang_mom}

The azimuthal component of the momentum equation is:
\begin{eqnarray}
\frac{1}{r^2} \frac{\partial}{\partial r} \left(
r^2  \rho  v_r  v_\phi \right)
&+& \frac {\partial}{\partial z} \left( \rho v_\phi v_z \right) \\
&=& \frac{1}{r^2}
\frac{\partial}{\partial r}\left( r^2 \langle t_{r \phi} \rangle  \right)  
+ \frac{\partial \langle t_{\phi z} \rangle}{\partial z} \qquad .
\nonumber
\label{e:azimuthal}
\end{eqnarray}
Integrating this equation  over $z$ and using the mass continuity equation (\ref{e:dM})
gives
\begin{equation}
\ddr \left[ \Mdota  v_\phi r + 2\pi r^2 T_{r\phi} \right]
= v_\phi r \frac{{\rm d}\Mdota}{{\rm d}r}
- 4\pi r^2 \langle t_{\phi z} \rangle^+   \qquad ,
\label{e:angmom}
\end{equation}
where
\begin{equation}
T_{r \phi} = \int_{-h}^{+h} \langle t_{r \phi} \rangle \> \dz
\label{e:int_stress}
\end{equation}
is the vertically integrated $r\phi$ stress. The terms on the the left
hand side of eqn.~(\ref{e:angmom}) describe radial transport of
angular momentum associated with radial inflow (accretion) and MHD
stresses acting over the disk height. The terms on
the right hand side describe vertical transport of angular momentum
resulting from mass loss in a wind and MHD stresses on the disk
surface. The magnetic part of the MHD stresses are given by
$\langle t_{r\phi} \rangle \sim \langle B_r
B_\phi \rangle/4\pi$ and  $\langle t_{\phi z} \rangle \sim \langle
B_\phi B_z \rangle/4\pi$ (cf eqn.~\ref{e:tij}) .

Radially integrating the angular momentum equation (\ref{e:angmom}) gives
\begin{eqnarray}
\Mdota   v_\phi   r 
- \Mdota (\ri)  v_\phi (\ri)  \ri =
-2\pi r^2 T_{r \phi} + 2\pi \ri^2 T_{r \phi}(\ri) \nonumber \\
  +
 \int_{\ri}^{r} \left[ \,  v_\phi r \frac{{\rm d}\Mdota}{{\rm d}r}
- 4 \pi r^2  \langle t_{\phi z}\rangle^+ \,  \right] \> {\rm d}r  \> .
\label{e:angmom_int}
\end{eqnarray}
This is a generalized conservation equation for angular momentum
in accretion disks. It can be equivalently expressed in terms of the
flux of angular momentum, $\dot J$, as follows:
\begin{equation}
\dot J_{\rm a}(r) - \dot J_{\rm a} (\ri) = \dot J_r (r) - \dot J_r
(\ri) + \dot J_z (r)
\qquad ,
\label{e:Jdot}
\end{equation}
where $\dot J_{\rm a} = \Mdota   v_\phi   r $ is the angular momentum
flux of the accreting matter, $\dot J_r = -2\pi r^2 T_{r \phi}$ is the
radial flux of angular momentum carried by the $r\phi$ stresses and $\dot
J_z = \int_{\ri}^{r} \left[ \,  v_\phi r \, {\rm d}\Mdota / {\rm d}r
- 4 \pi r^2  \langle t_{\phi z}\rangle^+ \,  \right] \; {\rm d}r$ is
the vertical flux of angular momentum carried by outflowing matter
and by the $\phi z$ stresses. Thus, both radial and vertical transport
contribute to the nett rate at which matter loses angular momentum and moves
radially inwards.  In other words, the mass accretion rate, $\Mdota$,
is determined by the nett rate at which angular momentum is transported
radially outwards by MHD stresses acting over the disk height and
vertically outwards by a mass outflow as well as by MHD stresses
acting over the disk surface.

The most luminous accretion-powered astrophysical
sources are inferred to be accreting matter at very high rates. Quasars, for
instance, appear to be accreting close to or even exceeding the Eddington rate,
$\Mdot_{\rm Edd} = 4\pi GMm_{\rm p}/(\eta \sigmaT c)
\approx 0.3 \, \eta_{0.1}^{-1} M_7 M_\odot\,{\rm yr}^{-1}$, where $\eta
= 0.1 \, \eta_{0.1}$ is the efficiency with which gravitational
binding energy is converted to radiation and $M =  10^7 M_7 M_\odot$
is the mass of the central black hole. Identifying the dominant
mechanisms responsible for such high mass accretion rates is a
major challenge confronting accretion disk theory.

It is interesting to compare the rate of angular momentum transport
indicated by (\ref{e:angmom_int}) with that predicted by standard
accretion disk theory \citep{SS73}. In a standard disk, angular
momentum is transported radially outwards only and the stresses
responsible for this are parameterized by a dimensionless parameter
$\alpha$ such that
\begin{equation}
T_{r \phi} = \alpha \cs h \Sigma r  \frac{\partial
  \Omega}{\partial r} 
\end{equation}
In addition, the stresses are assumed to vanish at $\ri$ and $\Mdota$
is necessarily constant with $r$ (since there is no mass
outflow). Thus, the angular momentum conservation equation,
(\ref{e:angmom_int}), reduces to
\begin{equation}
\Mdota r^2 \Omega = 2\pi \, r^3 \, \alpha \, \cs \, h \Sigma \,
\left| \frac{\partial \Omega}{\partial r}  \right|
\end{equation}
The parameter $\alpha$ lies in the range $0 \ltapprox \alpha \ltapprox
1$ for stresses resulting from subsonic turbulence. Values of $\alpha$
approaching unity are required to account for the
high accretion rates inferred in the most powerful sources, such as
quasars. However,
because the standard $\alpha-$disk model remains a phenomenological
prescription, it is unclear whether values $\alpha \sim 1$
correspond to a physically plausible realization of the
actual turbulent stresses in real accretion disks. 

Numerical simulations may be uniquely capable of addressing this
issue. Simulations of accretion disks have demonstrated conclusively
that turbulent MHD stresses can indeed remove angular momentum from
matter, thus facilitating the accretion process (see
\citealt{balbus03} for a review). However, 3D simulations of
MRI-generated MHD turbulence in accretion disks
have so far been unable to produce high mass accretion
rates \citep{hawley00,stonepring01,hawley01,hawbalb02}. Typical values of
$\alpha$ are $\mbox{a few}\times 10^{-2}$. On the other hand, high
accretion rates are recovered from 3D MHD simulations when a
large-scale, open mean magnetic field is explicitly
included \citep{steinhenn01,kigshib05}. These simulations and
others \citep[e.g.][]{salm07}, as well as semi-analytic models \citep[e.g.][]{campbell03},
show that at small disk radii, vertical transport of angular momentum by a large-scale
magnetic torque is more efficient than radial transport by MHD turbulence.

The numerical results indicate that MHD turbulence, whilst important
for the microphysics of accretion disks, cannot be solely responsible for
the removal of excess angular momentum from accreting matter. As
eqn.~(\ref{e:angmom_int}) shows, angular momentum can also be removed
by vertical components in the mean fluid and magnetic fields
(i.e. nonzero $v_z$ and $\langle B_\phi B_z \rangle$ components at the disk
surface -- see also \citealt{konpud00}). Large-scale MHD effects
in the form of mass outflows and magnetic
torques must therefore be primarily responsible for enhanced angular
momentum transport in accretion disks. The fundamental relationship
between angular momentum and energy (${\rm d} E = \Omega {\rm d} J$)
then also implies that large-scale MHD effects must also be
responsible for the high-energy phenomena associated with accreting
sources which must necessary originate outside the disk in a
relativistic jet or magnetized corona.

In the following section, we examine the global energy budget of
accretion disks in which angular momentum transport is prescribed by
the generalized angular momentum conservation equation (\ref{e:angmom_int}).

\subsection{Radiative disk flux}
\label{s:energy_budget}

We now consider energy conservation in MHD disk accretion. Again, we
refer the reader to our earlier paper \citep{kunbick04}
for details of the derivations.

Accretion power is the rate at which gravitational binding energy is extracted from the
accreting matter. This energy can be converted into mechanical
(e.g. kinetic, Poynting flux) and
non-mechanical (e.g. radiative) forms. The rate at which this
occurs is determined by keplerian shear in the bulk flow, $s_{r\phi} =
\frac{1}{2} r \partial \Omega / \partial r$, with $\partial \Omega /
\partial r = - \frac{3}{2} \Omega / r$. The rate per unit disk
surface area at which energy is emitted in the form of electromagnetic
radiation is determined by the internal energy equation. If there are
negligible changes in the internal energy and enthalpy of the gas,
then the disk radiative flux, $F_{\rm d}$, is approximately equal to
the rate of stochastic viscous dissipation of the turbulent energy,
which occurs on the smallest scales at the end of a turbulent
cascade. If there is negligible transport of turbulent energy from the
source region, then turbulent energy is locally dissipated at a rate
equivalent to its production rate, $\langle t_{ij} \rangle s_{ij}
\approx \langle t_{r\phi} \rangle s_{r\phi}$ (see \S~2.5 in
\citealt{kunbick04} for a discussion on the relative importance
of production, transport and dissipation of turbulent energy). The internal energy
equation then implies
\begin{equation}
\frac{\partial F_{\rm d}}{\partial z} \approx \langle t_{r\phi} \rangle s_{r
  \phi}
\end{equation}
Vertically integrating over the disk height yields the following
expression for the radiative flux emerging from the disk surface:
\begin{equation}
F_{\rm d}^+ \, \approx \, \frac{1}{2} T_{r\phi}  r  \frac{\partial
  \Omega}{\partial r}  \, = \, -\frac{3}{4} T_{r\phi} \Omega
\label{e:Fd_internal}
\end{equation}

The level of the turbulent MHD stresses $T_{r\phi}$ available for
internal dissipation depends on how efficient other processes are at
converting the extracted accretion energy into other (mechanical and
non-mechanical) forms. In other words, $T_{r\phi}$ is specified by the
angular momentum conservation relation (\ref{e:angmom_int}), which
gives
\begin{eqnarray}
-T_{r\phi} (r) &=& \frac{\Mdota v_\phi r}{2\pi r^2} \left[ 1 - \frac{\Mdota
    (\ri)}{\Mdota (r)} \left( \frac{\ri}{r} \right)^{1/2} \right]
    \nonumber \\
&-&
    \left( \frac{\ri}{r} \right)^2 T_{r\phi} (\ri)  \\
&-& \frac{1}{2\pi r^2}
\int_{\ri}^{r} \left[ \,  v_\phi r \frac{{\rm d}\Mdota}{{\rm d}r}
- 4 \pi r^2  \langle t_{\phi z}\rangle^+ \,  \right] \> {\rm d}r
\nonumber
\end{eqnarray}
Substituting this expression into (\ref{e:Fd_internal}) yields the
following solution for the disk radiative flux:
\begin{eqnarray}
\label{e:Fd}
F_{\rm d}^+ (r) &\approx& \frac{3GM\Mdota (r)}{8\pi r^3} \left[ 1 - \frac{\Mdota
    (\ri)}{\Mdota (r)} \left( \frac{\ri}{r} \right)^{1/2} \right]
    \nonumber \\
&-&
    \frac{3}{4} \left( \frac{\ri}{r} \right)^2 T_{r\phi} (\ri) \Omega
     \\
&-& \frac{3\Omega}{8\pi r^2} \int_{\ri}^{r} \left[ \,  v_\phi r \frac{{\rm d}\Mdota}{{\rm d}r}
- 4 \pi r^2  \langle t_{\phi z}\rangle^+ \,  \right] \> {\rm d}r \nonumber
\end{eqnarray}
The first two terms on the right hand side of this equation describe the
rate at which gravitational binding energy is extracted from matter as it accretes
(i.e. loses angular momentum); the second term in particular describes the rate at
which nonzero MHD stresses at the innermost stable orbit locally dissipate
turbulent energy (in practice, a convenient inner boundary condition is to set this
term equal to zero, although in principle, energy can still be
extracted beyond this boundary). The last term on the right hand side of
(\ref{e:Fd}) describes the rate at which energy is removed from
the disk by outflows involving mass and Poynting fluxes.

The result (\ref{e:Fd}) for the radiative flux of an accretion disk
corrected for the effects of outflows can be expressed as
\begin{equation}
F_{\rm d}^+ (r) \approx \frac{3GM\Mdota (r)}{8\pi r^3} \, \left[ \, f_{\rm
    a} (r) - \fw (r) \, \right] \qquad ,
\label{e:Fd1}
\end{equation}
where
\begin{equation}
\fa (r) = \left[ 1 - \frac{\Mdota
    (\ri)}{\Mdota (r)} \left( \frac{\ri}{r} \right)^{1/2} \right] -
\frac{2\pi \ri^2 T_{r\phi} (\ri)}{\Mdota (r) r^2 \Omega}
\label{e:fa}
\end{equation}
is the accretion factor and
\begin{equation}
\fw (r) = \frac{1}{\Mdota (r) r^2 \Omega} \int_{\ri}^{r} \left[ \,  v_\phi r \frac{{\rm d}\Mdota}{{\rm d}r}
- 4 \pi r^2  \langle t_{\phi z}\rangle^+ \,  \right] \> {\rm d}r
\label{e:fo}
\end{equation}
is the outflow correction factor (the `$\rm w$' subscript denotes
wind). Note that $\fw$ is equivalent to the fractional  rate of vertical angular
momentum transport in the disk. From
eqns.~(\ref{e:angmom_int}) and (\ref{e:Jdot}), we have
\begin{equation}
\fw(r) = \frac{\dot J_z (r)}{\dot J_{\rm a} (r)} \> .
\end{equation}
Similarly,
\begin{equation}
\fa(r)  = 1 - \frac{\dot J_{\rm a} (\ri) + \dot J_r (\ri)}{\dot  J_{\rm a}(r)} \> ,
\end{equation}
and hence, 
\begin{equation}
\fa(r) - \fw(r) = \frac{\dot J_r (r)}{\dot J_{\rm a}(r)} 
\end{equation}
Therefore, the efficacy of disk radiant emission is entirely determined by the
relative importance of radial transport of angular momentum in the
disk. If vertical transport of angular momentum dominates, then energy
is efficiently removed from the disk before being locally dissipated therein.

It is noteworthy to compare our outflow-modified disk flux with the
disk flux predicted by standard accretion disk theory.
In the absence of outflows
(that is, when ${\rm d}\Mdota /{\rm d} r =0$ and $\langle t_{\phi z}\rangle^+ =0$),
then $\fw = 0$, $\Mdota (\ri) = \Mdota (r) = \Mdot$ and hence, $f_{\rm
  a} = 1 - (\ri /r)^{1/2}$. In the limit of
vanishing stresses at $\ri$, this is equivalent to the small$-r$
correction factor from standard accretion disk theory \citep{SS73}. In
other words, standard accretion disk theory predicts that angular momentum
is transported radially outwards only and consequently, \textit{all}
the extracted gravitational binding energy is locally dissipated in
the form of radiation.


\section{Disk spectrum}
\label{s:spectrum}

An expression for the {\em total} disk radiative luminosity, $\Ld$, is
obtained by integrating each term in eqn.~(\ref{e:Fd}) over all disk
radii, from $r = \ri$ to $r = \infty$. Similarly, the spectrum of disk
emission can be calculated assuming local blackbody emission and
summing the spectrum from each annulus. This is the Multi-Colour Disk
(MCD) model used for standard disks \citep{mitsuda84}. We will henceforth
refer to our generalized disk prescription
as the Outflow-Modified Multi-Colour Disk (OMMCD) model.
 
Before we can calculate the total luminosity and emission spectrum for
the OMMCD model, it is necessary to define a specific functional form
for the outflows.
Realistically, of course, whether a disk becomes conducive to outflows
and the resulting behaviour of those outflows are determined by several
factors, including the degree of field-matter coupling, as well as the
field strength and topology, all of which will in general vary with
radius across the disk. Indeed, numerical simulations to date
have been unable to show whether a nonzero, nett poloidal field
can evolve from a stochastic field with zero nett flux. Here, we
wish to obtain a simple quantitative measure of how outflows might
affect the observable properties of an accretion disk. 

\subsection{A Simple Model}

A simple phenomenological model for
the mass accretion rate is  (see e.g. \citealt{wardkon93,li96,cassferr00}
for other self-similar models)
\begin{equation}
\Mdota (r) = \left\{ \begin{array}{ll}
\Mdot \left( \frac{r}{\rw} \right)^p & \> , \> r \leq \rw \\
\Mdot & \> , \> r \geq \rw
\end{array}  \right.
\end{equation}
where $\rw$ represents a critical wind radius beyond which the mass
outflow is negligible. The wind mass loss rate then satisfies $\Mdotw
(r) = \Mdot - \Mdota (r)$, from the continuity equation
(\ref{e:Mdot}). We can now use this model to explicitly calculate the
factor $\fa$, which appears as a source term in the disk flux solution
(\ref{e:Fd1}) and which is defined by (\ref{e:fa}). For the
simplest model, we assume that the stresses $T_{r \phi}(\ri)$ at the inner
boundary radius vanish, so that 
\begin{equation}
\fa (r) = \left\{ \begin{array}{ll}
1 - \left( \frac{r}{\ri} \right)^{-(1/2 +p)} & \> , \> r \leq \rw \\
1 - \left( \frac{r}{\ri}\right)^{-1/2} \left( \frac{\rw}{\ri}\right)^{-p} & \> , \> r \geq \rw
\end{array}  \right.
\label{e:fa_soln}
\end{equation}

Similarly, we can define a specific model for $\fw (r)$, given by
(\ref{e:fo}). From the
vertically-integrated, differential form of the angular momentum
equation, (\ref{e:angmom}), we have
\begin{eqnarray}
\label{e:out}
 r^2 \Omega \frac{{\rm d}\Mdota}{{\rm d}r}
&-& 4\pi r^2 \langle t_{\phi z} \rangle^+ \\
&=& 
\ddr \left( \Mdota r^2 \Omega \right)
 \left[ 1 + \frac{\ddr (2\pi r^2 T_{r\phi})}{\ddr (\Mdota r^2 \Omega)}
   \right]   \nonumber
\end{eqnarray}
We wish to define a model in which the relative importance of the outflows
in transporting angular momentum decreases with increasing radius. We choose a
simple power-law decline, such that
\begin{equation}
\left[ 1 + \frac{\ddr (2\pi r^2 T_{r\phi})}{\ddr (\Mdota r^2 \Omega)}
  \right] = \left( \frac{r}{\ri} \right)^{-w} \qquad ,
\label{e:w}
\end{equation}
with $w > 0$. Thus, in the limit $w \rightarrow 0$, radial
transport of angular momentum by the internal MHD stresses $T_{r\phi}$ is
negligible and the dominant mode of transport is via outflows.
Inserting the relation (\ref{e:w}) into (\ref{e:out}) and simplifying, we have
\begin{eqnarray}
v_\phi r \frac{{\rm d}\Mdota}{{\rm d}r}
&-& 4\pi r^2 \langle t_{\phi z} \rangle^+ \\
&=& 
\left( p+\frac{1}{2} \right) \Mdot c \left( \frac{\rg}{\ri}
\right)^{1/2} \left( \frac{\ri}{\rw} \right)^p \left( \frac{r}{\ri}
\right)^{-(1/2-p+w)}
\nonumber
\end{eqnarray}
This implies the following relation for the outflow factor 
$\fw$ defined by (\ref{e:fo}):
\begin{eqnarray}
\fw (r \leq \rw) &=& \left( \frac{1/2+p}{1/2+p-w} \right) \left(
\frac{r}{\ri} \right)^{-w} \nonumber \\
&\times&\left[ 1 - \left( \frac{r}{\ri} \right)^{-(1/2+p-w)}  \right]
\end{eqnarray}
\begin{eqnarray}
\fw (r \geq \rw) &=& \left( \frac{1/2+p}{1/2+p-w} \right)
\left( \frac{r}{\rw} \right)^{-1/2}
\left( \frac{\rw}{\ri} \right)^{-w} \nonumber \\
&\times& \left[ 1 - \left( \frac{\rw}{\ri} \right)^{-(1/2+p-w)} \right] \nonumber  \\
&+& \left( \frac{1/2}{1/2-w}\right) \left( \frac{r}{\ri} \right)^{-w}
 \\
&\times& \left[ 1 - \left( \frac{r}{\rw} \right)^{-(1/2-w)} 
\right] \nonumber
\label{e:fw_soln}
\end{eqnarray}

Eqns.~(\ref{e:fa_soln}) and (\ref{e:fw_soln}) for $\fa$ and $\fw$,
respectively, now specify the solution for the OMMCD radiative flux as a function
of radius, given by eqn.~(\ref{e:Fd}).

The total disk luminosity can now be
calculated by integrating the flux over all disk annuli, as follows:
\begin{eqnarray}
\Ld &=& 2 \int_{\ri}^\infty \Fd (r) \, 2\pi r \, \dr \\
 &=& \frac{3}{2} \frac{GM \Mdot}{\ri} \int_{\ri}^\infty \frac{\Mdota
 (r)}{\Mdot} \left( \frac{r}{\ri} \right)^{-2} \left[ \,
 \fa(r) - \fw (r) \, \right]  \, \frac{\dr}{\ri} \nonumber 
\label{e:Ld}
\end{eqnarray}
We can write the solution as
\begin{equation}
\Ld \, = \, \Pa \, - \, \Pw \qquad ,
\end{equation}
where
\begin{eqnarray}
\Pa = \frac{1}{2} \frac{GM \Mdot}{\ri}
\left( \frac{1+2p}{1-p} \right) \left( \frac{\rw}{\ri} \right)^{-p}
\nonumber \\
\left[ 1 - \frac{3p}{(1+2p)} 
\left( \frac{\rw}{\ri}  \right)^{-(1-p)}
\right]
\label{e:Pa}
\end{eqnarray}
is the total accretion power and
\begin{eqnarray}
\Pw = \frac{1}{2} \frac{GM \Mdot}{\ri}
\left( \frac{1+2p}{1-p+w} \right) \left( \frac{\rw}{\ri} \right)^{-p}
\nonumber \\
\left[ 1 - \frac{p(3+2w)}{(1+2p)(1+w)}
\left( \frac{\rw}{\ri} \right)^{-(1-p+w)}
\right]
\label{e:Pw}
\end{eqnarray}
is the total power removed by outflows. Note that in the limit $p
\rightarrow 0$, corresponding to negligible mass loss (i.e. 
$\Mdota \rightarrow constant$), the accretion power approaches the
solution from standard accretion disk theory, $\Pa = GM\Mdota/2\ri$,
and the wind power is $\Pw \rightarrow \Pa/(1+w)$. Thus, a substantial fraction
of the total accretion power can be removed from the disk by a magnetic torque
alone. This solution represents a Poynting flux dominated outflow and
can be identified with the formation of
relativistic jets that carry away very little matter, but transport a large
amount of mechanical energy very efficiently across vast distances.

The emission spectrum predicted by the OMMCD model can be calculated in
the same way as that for an MCD model (i.e. a standard accretion
disk). Assuming each annulus in the disk radiates locally like a
blackbody, $B_\nu$, the total disk spectrum is calculated by summing
the individual spectra from each annulus:
\begin{equation}
L_{\rm d , \nu} = 2 \int_{\ri}^\infty \pi B_\nu [ T(r) ] \, 2\pi r \,
\dr
\qquad ,
\end{equation}
where $T(r) = [ \Fd (r) /\sigma ]^{1/4}$ is the effective disk
temperature of each annulus and $\sigma$ is the Stefan-Boltzmann constant.

\begin{figure}[ht!]
\centerline{ \includegraphics[width=0.53\textwidth]{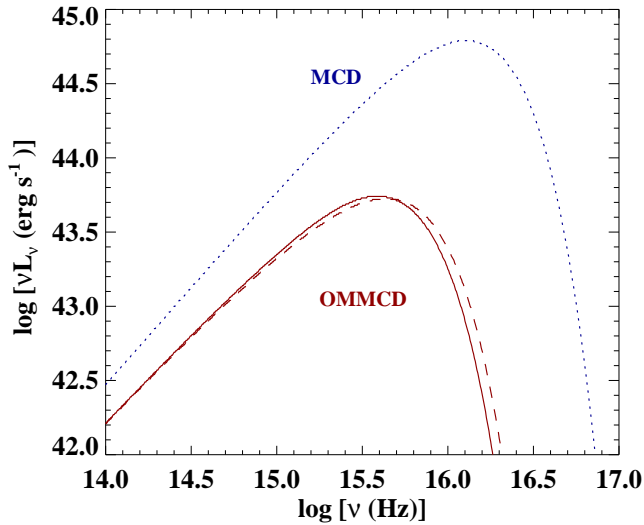}}
\caption{Comparison of theoretical accretion disk spectra . The MCD
  model (dotted blue curve) is based on standard accretion disk
  theory. The OMMCD model (red curves) predicts a radial disk structure modified by
  outflows. The dashed red curve is the predicted OMMCD spectrum for a
  Poynting flux dominated outflow, while the solid red curve is the
  OMMCD spectrum for a mass-loaded outflow. In all models, the black
  hole mass is $M = 10^7 M_\odot$ and the mass accretion rate at
  infinity is $\Mdot = 0.27 \, M_\odot \, {\rm yr}^{-1}$.}
\label{nuLnu}       
\end{figure}

Figure~\ref{nuLnu} shows theoretical disk emission spectra $L_{\rm d
  ,\nu}$ for three different models of black hole disk accretion:
  1. the MCD model (blue dotted curve) derived from the standard ``Shakura-Sunyaev''
  theory \citep{SS73}; 2. the OMMCD model with negligible
mass loss in an MHD outflow ($p=0.1$, $w=0.1$, red dashed curve); and 3. an
  OMMCD model with stronger mass loss in an MHD outflow
($p=0.5$, $w=0.1$, red solid curve). In both OMMCD models, the mechanical
  energy carried away by the outflow is
$\approx 90\,$\% of the total accretion power and mass
loss is restricted to radii $r \leq 100\rg$. In all three models, $M=
  10^7 M_\odot$ (corresponding to $\LEdd \approx 1.3 \times 10^{45}
  {\rm erg \, s}^{-1}$) and $\Mdot = 0.27 \, M_\odot \, {\rm yr}^{-1}$. This
  is the  nett mass influx at $r=\infty$ and corresponds to the Eddington
  rate when the radiative efficiency is $\eta \approx 0.08$. This is
  strictly only the case for the MCD disk model. When outflows are
  important,  the radiative efficiency can fall below the nominal
  $8$\% predicted by standard theory because the outflows remove
  energy from the disk (as well as angular momentum), so there is
  relatively less energy to dissipate and radiate away. For the OMMCD
  models shown in Fig.~\ref{nuLnu}, the disk
  radiative efficiency is about
  an order of magnitude lower than that of a standard disk.

As is clearly evident in Fig.~\ref{nuLnu}, outflows can substantially
modify the overall spectrum of emission. The generalized OMMCD model
predicts a high-energy cut-off in the emission spectrum at lower
energies than that predicted by the standard MCD model. This is a
direct results of the importance of outflows at small radii, where the highest-energy
emission originates. In the case of stellar-mass black holes, the
spectrum is most affected at X-ray energies (around a few keV), while
for supermassive black holes, it is the extreme ultraviolet region
($\sim 10^{16}\,{\rm Hz}$) that is most affected by outflows. The
OMMCD model also predicts a broadband region of the disk spectrum
(typically around visual wavelengths) that is much flatter than the
characteristic $\nu^{1/3}$ law predicted by standard theory.

We have used a specific phenomenological model to quantitatively
calculate how a disk spectrum can be modified by outflows and a different choice for
the functional form of the outflows may result in a somewhat different predicted
spectral shape. Nevertheless, the overall effect of outflows on a disk should be
largely insensitive to the detailed physical properties of the
outflows; the extraction of angular momentum and binding energy from
the disk will inevitably result in an emission spectrum that is dimmer
and redder than that predicted by the standard model for the same mass
accretion rate.

\subsection{Comparison with observations}

Our results demonstrate that the modifications to a standard disk spectrum as a result of energetic outflow phenomena are clearly not negligible when the outflows are primarily responsible for the removal of angular momentum from accreting matter. Direct observational verification of the predicted OMMCD spectrum may be possible for only some types of accreting sources, however, as the outflows will also produce their own characteristic radiative signatures that may overlap in the spectral energy band where we predict the disk to be most strongly modified. This can confuse our interpretation of the emergent observed spectrum.

In the case of galactic X-ray binaries (XRBs), for instance, the disk spectrum can be substantially modified at X-ray energies (typically around $1$\,keV). These sources often exhibit a power-law X-ray spectrum above $1$\,keV that is thought to arise in a magnetized, tenuous corona and/or relativistic jet as a result of inverse Compton scattering off disk photons (see \citealt{mcclinrem06} for a review). Existing models for this emission process simply use a standard disk spectrum for the seed photon distribution. However, this is clearly not a self-consistent calculation since the formation of a corona and/or jet from disk magnetic fields inevitably modifies the disk radial structure and hence,  photon spectrum, as we have shown here. Thus, although we would not expect to directly see the modified disk spectrum in XRBs, our model can be indirectly tested by using the underlying OMMCD spectrum as the source of seed photons that are upscattered in a disk corona and/or jet. By comparing the predicted X-ray spectrum against the observed spectrum, we can improve our interpretation of the source characteristics and obtain tighter constraints on key physical parameters such as mass accretion rate.

The above type of calculation has been performed by us for ultra-luminous X-ray sources (ULXs) \citep{freeland06}. These exceptionally luminous XRBs probably involve accretion onto a black hole more massive than those found in galactic XRBs. They provide an effective test for our modified disk model because some ULXs exhibit a low energy ($\ltapprox 1\,{\rm keV}$) spectral component that can be interpreted as disk emission. However, because these sources emit virtually all of their radiative power as hard, power-law X-rays, which must necessarily be produced outside the disk, the accretion disk must be substantially modified and a standard disk model should not be used to fit the soft spectral component. Indeed, we show \citep{kuncic06} that the OMMCD spectrum can adequately fit the observed soft spectral component, implying a black hole mass $M \sim 100 M_\odot$.

A similar test can also be performed for quasars and other AGN,
powered by accretion onto a supermassive ($10^{6-9}\,M_\odot$) black
hole. The prominent optical/UV continuum feature known as the "big
blue bump" \citep{sanders89} is generally interpreted as accretion
disk emission \citep{shields78,malksarg82}. However, attempts at
fitting accretion disk spectral models to the observed spectra have had mixed success
(see \citealt{koratblaes99} for a review). For AGN, the disk spectrum
will be most strongly modified by outflows at UV wavelengths (see
Fig.~\ref{nuLnu}). Although there is less overlap between the modified
disk and corona/jet X-ray emission than in the XRB case, the poor
transmission of UV radiation through Earth's atmosphere means that
direct observations of the predicted OMMCD spectrum in AGN are not
straightforward. Nevertheless, spectral energy distributions (SEDs)
have been compiled from observations of high-redshift quasars with the
SDSS \citep{richards06,trammell07} and
from satellite UV observations with the \textit{HST}-FOS
\citep{zheng97,telfer02}
and with \textit{FUSE} \citep{scott04,shang05}. Interestingly, these
observations have revealed a
far-UV break in the SEDs of quasars (see \citealt{trammell07} and
references cited therein). Specifically, the observed spectra decline
dramatically at wavelengths shorter than $1100$\,\AA\,
(i.e. frequencies above $10^{15.4}\,{\rm Hz}$). This is
difficult to explain with standard accretion disk models, which
predict that the spectrum should continue to rise into the far and extreme UV
(see Fig.~\ref{nuLnu}).

\begin{figure}[ht!]
\centerline{ \includegraphics[width=0.53\textwidth]{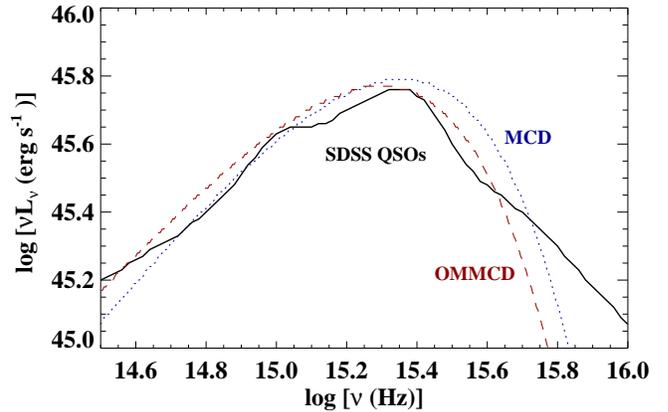}}
\caption{A comparison between observed and predicted quasar optical/UV
luminosity spectra. The solid black curve is the restframe mean
spectral energy distribution of SDSS type I quasars, corrected for galactic
extinction (see Richards et al. 2006 for online data). The
dotted blue curve is the spectrum predicted by a standard multi-colour
disk (MCD) model with $M = 10^9 M_\odot$ and $\Mdota = 2.7 \, M_\odot \, {\rm yr}^{-1}$
(equivalent to $\Mdota = 0.1 \, \dot M_{\rm  Edd}$). The dashed red curve
is the spectrum predicted by the outflow modified multi-colour disk
(OMMCD) model, with $M = 5 \times 10^8 M_\odot$ and $\Mdot = 25 \, M_\odot
\, {\rm yr}^{-1}$ (corresponding to $\Mdota (\infty) = 19 \, \dot M_{\rm
  Edd}$), where 90\% of the accretion power is carried away by an MHD outflow.}
\label{nuLnu_SDSS}       
\end{figure}

Figure~\ref{nuLnu_SDSS} shows a comparison of the MCD and OMMCD
spectra against the mean SED of SDSS type I quasars
\citep{richards06}. The parameters used for the MCD spectrum (dotted
blue curve) are:
$M = 1 \times 10^9 M_\odot$ and $\Mdot = \Mdota = 2.7 \, M_\odot \, {\rm
  yr}^{-1}$, which is equivalent to $0.1 \dot M_{\rm Edd}$. The OMMCD
spectrum uses $M = 5 \times 10^8 M_\odot$ and $\Mdot = 25 \, M_\odot \, {\rm
  yr}^{-1}$, which corresponds to a mass accretion rate of about $19
\dot M_{\rm Edd}$ at $r=\infty$. The outflow parameters are $p=0.5$,
$w=0.1$ and $\rw / \ri =10$. In this case, only 30\% of the matter accreting at
$r=\infty$ reaches $\ri$ and 90\% of the total accretion power is
removed by the outflow. Because most of this energy is removed
preferentially at small radii, the disk is not as hot at these radii
as a standard disk. Consequently, the emergent disk spectrum is fainter
in the far-UV, precisely in the bandpass where a break has been detected
independently by several instruments. Note, however, that both the MCD and OMMCD models underpredict what appears to be a secondary spectral component in the extreme UV (at frequencies above $10^{15.7}\, {\rm Hz}$). This component resembles a power-law with a spectral index close to $2$ and could result from Comptonization.

Fig.~\ref{nuLnu_SDSS} clearly shows the
difference between the far-to-extreme UV spectrum predicted by a standard disk and
that predicted by an outflow modified disk. There are two important
points to draw from this comparison. Firstly, the standard
model does not take into account the partitioning of accretion
energy into disk radiation and other non-radiative forms of energy
(e.g. Poynting flux). The second important point is that the two
different disk models predict different values for the fundamental
physical parameters $M$ and $\Mdot$. The implication is then that we
could be severely misinterpreting quasars and AGN by using the
standard disk model to infer, in particular, mass accretion rates. The
universality of accretion related outflow phenomena suggests that the
true rates of mass accretion could be considerably higher than we are
currently estimating them to be.

\section{Concluding Remarks}

The standard theory for astrophysical disk accretion has enjoyed
remarkable success since its inception over 30 years ago. However,
ongoing rapid advances in instrumental technology are placing
increasingly tighter observational constraints on interpretational models
that utilize the standard theory. The ubiquity of energetic outflow
phenomena in accreting astrophysical sources gives us two important
clues to the underlying processes responsible for accretion:
(1.) Vertical transport of angular momentum cannot be ignored and may
indeed be the dominant mode of transport at small radii; and (2.) The
removal of accretion energy from the disk results in a modified
radial disk structure at small radii and  a
disk spectrum that can be substantially different from that predicted by
standard accretion disk theory.

We have presented a simple model that can be used
to quantitatively calculate model spectra for disks modified by a
magnetized jet and/or corona and by mass-loaded winds. These model
spectra can be applied to a variety of different sources to improve
our interpetation of the observations and more accurately determine
key physical parameters such as mass accretion rate and black hole
mass, as well as the partitioning of accretion power into radiative
and non-radiative forms.

%

\begin{acknowledgements}
The authors wish to thank all the particpants for contributing to a
very successful and very informative Fifth Stromlo Symposium.
\end{acknowledgements}




\end{document}